# Vertical Heterostructures between Transition-Metal Dichalcogenides - A Theoretical Analysis of the NbS2/WSe2 junction

Zahra Golsanamlou[1], Poonam Kumari[1], Luca Sementa[1,*], Teresa Cusati[2], Giuseppe Iannaccone[2], Alessandro Fortunelli[1]*

[1] CNR-ICCOM and IPCF, Consiglio Nazionale delle Ricerche, via G. Moruzzi 1, Pisa 56124, Italy

[2] Dipartimento di Ingegneria dell'Informazione, Università di Pisa, Via G. Caruso 16, Pisa 56122, Italy

* Corresponding authors: luca.sementa@cnr.it, alessandro.fortunelli@cnr.it

ORCID AF = 0000-0001-5337-4450, ZG= 0000-0002-6316-4147, LS = 0000-0002-3951-2842

**Abstract**

Low-dimensional metal-semiconductor vertical heterostructures (VH) are promising candidates in the search of electronic devices at the extreme limits of miniaturization. Within this line of research, here we present a theoretical/computational study of the $NbS_2/WSe_2$ metal-semiconductor vertical hetero-junction using density functional theory (DFT) and conductance simulations. We first construct atomistic models of the $NbS_2/WSe_2$ VH considering all the five possible stacking orientations at the interface, and we conduct DFT and quantum-mechanical (QM) scattering simulations to obtain information on band structure and transmission coefficients. We then carry out an analysis of the QM results in terms of electrostatic potential, fragment decomposition, and band alignment. The behavior of transmission expected from this analysis is in excellent agreement with, and thus fully rationalizes, the DFT results, and the peculiar double-peak profile of transmission. Finally, we use maximally localized Wannier functions, projected density of states (PDOS), and a simple analytic formula to predict and explain quantitatively the differences in transport in the case of epitaxial misorientation. Within the class of Transition-Metal Dichalcogenide systems, the $NbS_2/WSe_2$ vertical heterostructure exhibits a wide interval of finite transmission and a double-peak profile, features that could be exploited in applications.

**Introduction**

Two-dimensional (2D) transition metal dichacogenides (TMDs) have attracted extensive attention in recent years due to their promising applications in electronic and optoelectronic devices, [1, 2, 3, 4, 5]. In particular, their lateral hetero-structures (LH) and vertical hetero-structures (VH) may allow realizing metal/semiconductor junctions at the ultimate thickness limit, that could be competitive, at the fundamental level of control of the electrostatic potential in the device, with present silicon-based integrated circuit technology [3,4, 6]. Device engineering has been shown to play an important role in improving performance of these systems, since issues associated with lattice mistmatch and Fermi level pinning in metal-semiconductor devices, and (as we will see later) the presence of a pseudo gap below the Fermi level for these non-ideal metals can notably affect charge transport [3, 7, 8, 9]. Experiments have indicated that interfaces based on van der Waals (vdW) forces as in VH instead of covalent ones as in LH can help decrease some of these issues. In VH the hetero-junction is realized by stacking one (few) layers of one TMD on top of one (few) layers of a different TMD, and the junction is held together by vdW forces, whereas in LHs the stoichiometry changes from one TMD to another within the same single layer, and the two TMD pieces are held together by covalent/ionic forces. The presence of weaker (dispersion) forces in VH junctions arguably lead to a decrease of the gradient of the electrostatic potential at the interface, as well as reducing the probability of chemical disorder and defect-induced gap states which often occur in LHs (in addition to band engineering effects as we will see later), therefore explaining the typically superior performance of VH-based devices. As a result, vertically stacked heterostructures are a subject of considerable attention for 2D TMD devices [3, 8, 10, 11].

At the experimental level, Shin et al. investigated charge transport in $NbS_2$/n(electron acceptor)-$MoS_2$ VH [11]. They calculated the work function (WF) of $NbS_2$ and n-$MoS_2$ and measured the associated Schottky barrier, showing that this barrier affects charge transfer and causes carrier depletion in $MoS_2$ underneath $NbS_2$. Also, they showed that the vdW interface between $NbS_2$ and $MoS_2$ favors in-plane carrier transport, and that charge transport can be tuned by applying bias and gate voltage.  p-n (electron donor-electron acceptor, respectively) VHs composed of two semiconductor TMDs have also been widely studied at both experimental and theoretical level [12, 13, 14, 15, 16, 17]. Huo et al. studied the transport properties of the $WS_2$/$MoS_2$ VH [15] via band structure calculations, showing that the bottom of the conduction band (BCB) and the -top of the valence band (TVB)  of $WS_2$ are higher in energy than those of $MoS_2$, so that electrons are transferred from $WS_2$ into $MoS_2$. Lee et al. investigated the optoelectronic properties of the $WSe_2$/$MoS_2$ VH [16]. They reported that charge transport occurs mostly in the vertical direction due to the voltage drop at the interface of the p-n junction, without a significant potential barrier. Therefore, charge transfer occurs via interlayer tunneling of majority carriers between the BCB of $MoS_2$ and the TVB of $WSe_2$. Band alignment of $MoS_2$/$MoTe_2$ vdW heterojunction on a $SiO_2$ substrate was calculated by Lim et al. using data extracted from scanning Kelvin probe microscopy (SKPM) [17]. The work functions of the individual $MoTe_2$ and $MoS_2$ phases turned out to be identical, implying that MoS$_2$ has a slightly higher work function in the heterojunction. This change in work function was assumed to be due to charge transfer between the two TMDs and also to trap charges in the SiO$_2$ substrate. They also used gate voltage and bias to change the type and rate of charge transfer. Without a gate bias, they showed that the junction between the MoS$_2$ n-channel and the MoTe$_2$ p-channel has almost no barrier, but there is a built-in potential for the MoS$_2$ n-channel due to the difference in work functions. Transport properties of vertical semiconductor/metal and metal/metal heterojunctions made of TMDCs (MoS$_2$/NbS$_2$, MoSe$_2$/NbS$_2$, NbS$_2$/NbSe$_2$ and CoS$_2$/CoSe$_2$) along with their electronic properties were studied by Costa et al. [18]. They confirmed that, because of the large work function of the metals (NbS$_2$ and CoS$_2$), the TVB of the semiconductor is typically positioned above the Fermi energy of the metal. They also discussed how the alignment of the projected density of states (PDOS) in the metal and semiconductor phases and the pinning of the TVB of the latter to the Fermi level of the former makes that the junction show a diode-like behavior.

We underline that the phenomenon by which the TVB of the semiconductor is pinned at the Fermi energy of the metal is of crucial importance in these systems (as we will see hereafter), and in this respect these 2D materials behave almost exactly as their 3D counterparts. In general, the tunneling barrier at the interface of metal-semiconductor junction has thus typically a Schottky character. There are different ways in literature to calculate the Schottky barrier height (SBH). Jelver et al. analyzed the slope of density of states (DOS) and the transmission values to estimate the SBH [19]. They also showed that the presence of localized states due to the interface or to doping can modulate the barrier height. In previous work, we estimated the SBH from the jump in electrostatic potential at the interface [20], i.e., by the difference between the local Fermi energy of the metallic part and the local TBV of the semiconductor part precisely at the interface, where a fragment-analysis method was used to obtain the needed alignment of the two components [21]. The origin and nature of Schottky barriers in TMD monolayers (MX$_2$ where M = Mo, W and X = S, Se, Te) in contact with an ideal metal electrode have been studied by D. Szcz niak et al. [22] using combination of the complex band structure formalism and the cell-averaged Green's function technique. They found the charge neutrality level (CNL) and calculated the SBH as the difference between the CNL and the metal work function. They noted that the CNL lie at the midpoint of the semiconducting bandgap, and attributed the origin of SBHs to metal induced gap states (MIGs), analogously to the case of 3D metal-semiconductor junctions. The SBHs of two dimensional TMD(MoS$_2$, MOSe$_2$, MOTe$_2$)/metal(Sc, Mg, Al, Ti, Cr, Mo, Ru, Co, Ni, Pd, Pt, MoO$_3$) junctions for both top and edge contact geometries have also been studied by Y. Guo et al. [23], showing that for both the edge and top contact geometries, SBHs are pinned to the Fermi energy of the metal with similar pinning factor. This was attributed to the direct bonding between the contact metal atom and TMD chalcogen atom in both cases.

Using density-functional theory (DFT) Ren et al. considered the different stacking of GeC on TMD$_2$ (MoS$_2$, MoSe$_2$, WS$_2$ and WSe$_2$) and studied electronic structure and electrostatic potential for the most stable vdW geometries [24]. An atom-projected band structure analysis of the heterostructure showed that the TMDs and GeC preserve their direct bandgaps, and the bandgap of each studied heterostructure is less than the band gap of the constituent systems. The analysis of band alignment indicates electron transfer from TMDs to GeC which causes a potential drop and a built-in electric field across the vdW heterostructure. Cusati et al. investigated the effect of the stacking sequence on vertical electron transport in $MoS_2$ multilayer systems [25] and concluded that interlayer electron transport is strongly dependent on the stacking of the $MoS_2$ layers, and also that the interlayer distance (related to mis-orientation phenomena) also affects the transmission coefficient. Work function engineering for metal/semiconductor VHs was investigated by Ding et al. [26, 27]. They considered 2D H(phase)- and T(phase)-$MX_2$ (M = Ti, V, Nb, Ta, Mo, and W; X = S and Se) as metals and H-$WSe_2$ as semiconductor, respectively. They extracted the value of the SBH from an analysis of band edge position, investigated the behavior of the SBH as a function of the work function of the different metals, and quantified the strength of pinning in terms of the derivative of the SBH with respect to the metal work function. They found that the ideal Schottky–Mott limit is approached when the SBH changes exactly in parallel with the work function as occurring in the region of low values of the metal work function, whereas a strong pinning-like behavior is found to occur in the Ohmic contact region of high values of the metal work function. Hu et al. studied the electronic structure, band alignment and potential barrier for the 1T(phase)$VSe_2$/multilayer-MoSSe VH, together with other Schottky systems for comparison, including $WSe_2$ [8], in the search for Ohmic contacts. They found that Schottky and tunnel barriers of 1T(phase)$VSe_2$/multilayer-MoSSe VH can be regulated by changing the vdW stacking at the interface, or by applying a tensile strain: by decreasing the BCB and increasing the TVB of $VSe_2$, the Schottky-to-Ohmic contact transition can be achieved.

Here we study the $NbS_2$/$WSe_2$ 2H-bulk VH as a prototype of metal-semiconductor VH hybrid interface. After building realistic structural models, we consider the 5 possible orientations of the layers at the interface (5 different stackings), derive their band structure, and calculate transmission coeffcients in these systems, finding a non-zero interval wider than in the corresponding LH and an interesting *double-peak profile* of transmission. We then apply an analysis of electron transport via two tools: (1) fragment decomposition and electrostatic potential analyses, and (2) alignment of the PDOS of the atoms in the scattering region. With these tools, we are able to fully rationalize the behavior of the transmission coefficient and its double-peak profile in the case of minimal-distance interlayer epitaxy, tracing back the double peak profile ultimately to the presence of a pseudo-gap below the Fermi energy in $NbS_2$, as is typical in the DOS of these non-ideal metallic layered materials. However, we find that band structure and alignment analysis is not able to explain transport when the stacking distance at the interface is increased to that expected for mis-oriented (orientationally mis-matched) stackings. We then introduce a simple model combining atomic PDOS and hopping elements at the interface as derived from a Wannier analysis of the wave function, and show that this model can now accurately explain the computational results, in particular the presence or absence of a double-peak profile in transmission depending on interfacial distances. Also, in comparison with the homologous $WSe_2$/$NbS_2$ lateral heterostructure [20], the wide-interval and double-peak-profile transmission here found for the VH system is of interest for potential applications, while the theoretical fragment, alignment, and Wannier analysis of electronic transmission should be generally applicable to the study of transport in 2D materials.

Band alignment in VH also depends on the local pressure. A change in pressure will change interlayer distances which in turn will affect the tight-binding hopping matrix elements as well as the DOS, so that one expects significant changes in the SBH as well as in transmission as a function of pressure. Indeed, de Araújo et al. studied the change in Schottky barrier as a function of applied force (pressure) in Ref.[28], finding that SBH changes at a rate of 0.21(for a few layers $MoS_2$), 0.23 (for three layers $MoS_2$), and 0.78 (for two layers $MoS_2$) meV $nN^{-1}$.

The article is organized as follows. The computational approach is presented in Section 2. Section 3 presents and discusses our main results, while conclusions are summarized in Section 4.

## 2. Methods

Electronic structure calculations were performed using Quantum Espresso (QE) package [29, 30], using 50 Ry as energy cutoff for the wave function and 500 Ry as density cutoff for the density, and large 22×22×1 Monkhorst-Pack k-meshes to sample the Brillouin zone (BZ). A plane-wave basis set, a gradient-corrected exchange-correlation (xc-)functional [the Perdew-Burke-Ernzerhof (PBE)] [31] and scalar-relativistic ultrasoft pseudopotentials (US-PPs) were utilized.

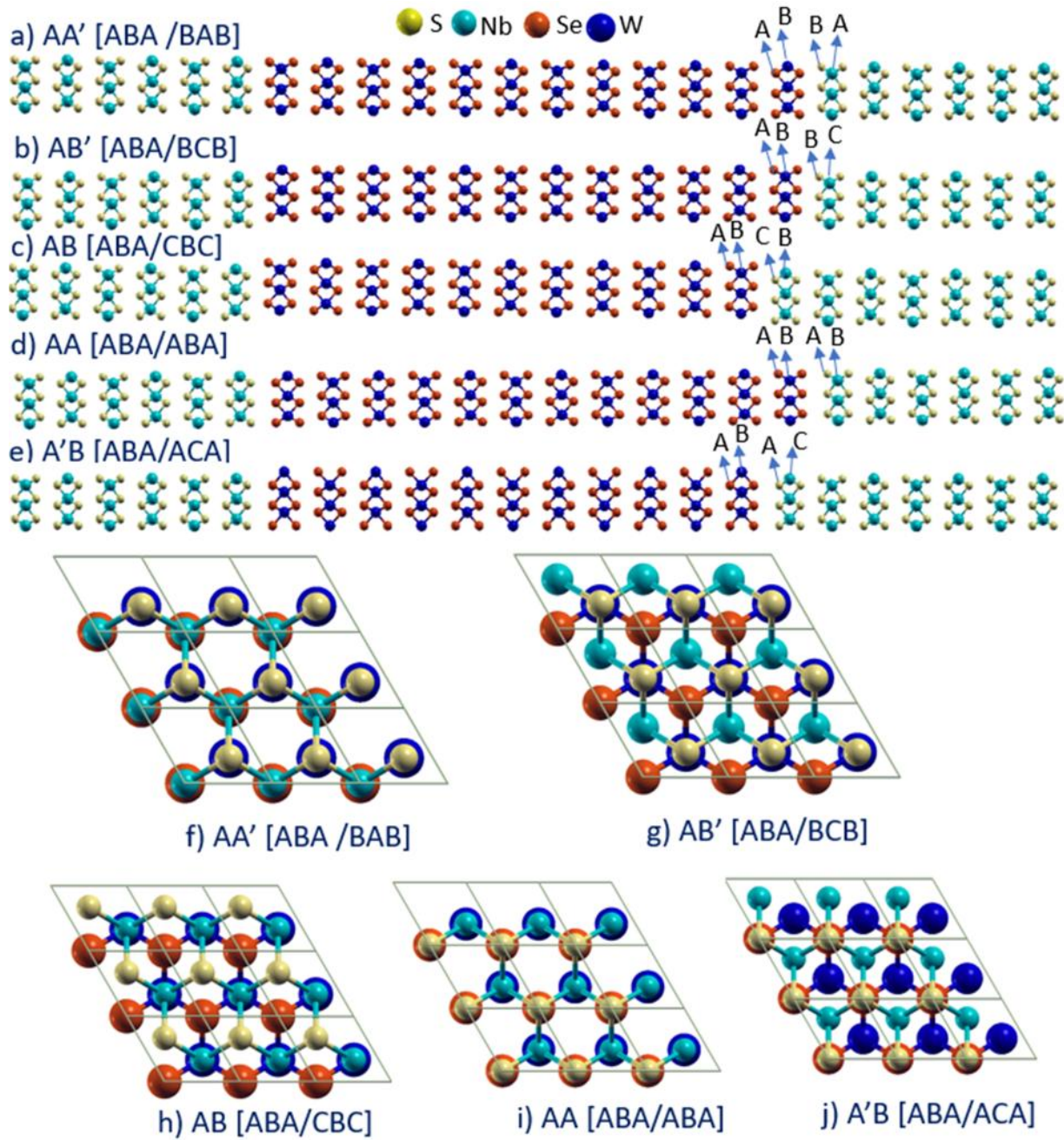

Figure 1. The five stacking geometries considered for the NbS$_2$/WSe$_2$ VH, representing Nb atoms are in cyan color, S atoms are in yellow, W atoms are in dark blue and Se atoms are in dark orange. (a-e) side views: a) AA'[ABA/BAB] in the left interface TM atom from last NbS$_2$ layer goes on top of X (chalcogen atom) of the first WSe$_2$ layer – in the right interface the TM atom of the last WSe$_2$ layer goes on top of the X atom of the first NbS$_2$ layer , b) AB'[ABA/BCB] in the left interface the TM atom of last NbS2 layer goes on the top of the X atom of first WSe2 layer while the X atom from the NbS2 layer goes in the hollow position of the WSe2 layer,  c) AB[ABA/CBC] in the left interface the TM atom from the last NbS$_2$ layer goes on top of the TM atom of the first WSe$_2$ layer, while the X atoms from the WSe$_2$ layer sit in the hollow position of the NbS$_2$ layer,  d) AA [ABA/ABA], in the left interface the TM and X atoms from the last NbS$_2$ layer are stacked over the TM and X atoms of the first WSe$_2$ layer – in the right interface the TM and X atoms of last WSe$_2$ layer stacked over TM and X atoms of first NbS$_2$ layer ,  and e) A'B [ABA/ACA], in the left interface the X atom from last NbS$_2$ layer goes on top of X atom from the first WSe$_2$ layer, while the TM atoms from one layer at the interface sits on the hollow position of the next layer. In the right interface the X atom from last WSe$_2$ layer goes on top of the

*X atom from the first NbS2 layer, while the TM atoms of the NbS2 layer sit in a hollow position of WSe2 layer. (f-l) top views of the structures which shows the right interface of the structure explicitly: f) AA', g) AB', h) AB, i) AA and l) A'B*

Transmission simulations were carried out based on a scattering state approach considering rightward propagating modes from the left to the right electrode as implemented in the QE/PWcond routine [32, 33] in the limit of ballistic transport. In the transmission simulations we used the same (22×22×1) k-mesh as in QE/PWscf (see Supplementary Information, Section 1). The DFT band structure was then analyzed to derive Hamiltonians based on maximally localized Wannier functions via the Wannier90 code [34, 35], using transition-metal *d*-orbitals and chalcogenide *p*-orbitals in the basis set. More details about Wannier90 are given in the Supplementary Information.

Five stacking geometries are possible in a VH between two transition metal dichalchogenide layered systems [8,25], conventionally denoted as: AA', AB', AB, AA, A'B, as depicted in Figure 1. Using a more illustrative and physically clearer notation, focusing on the atoms of the two monolayers at the interface, AA' can be denoted as ABA/BAB, where A,B,C are the conventional positions in (111) stacking and correspond, in order, to the positions of: (first S atoms in $NbS_2$)(Nb in $NbS_2$)(second S atoms in $NbS_2$)/(first Se atoms in $WSe_2$)(W in $WSe_2$)(second Se atoms in $WSe_2$). Here we considered all the 5 possible stackings between $NbS_2$ and $WSe_2$ phases assumed in 2H bulk arrangement, i.e.:

a) AA': the stacking of atoms at the interface can be described as ABA/BAB, thus S/Se first-neighbors are in *hollow positions*, the S atoms eclipse W atoms, and the Se atoms eclipse Nb atoms – AA' corresponds to the bulk 2H stacking;

b) AB': the stacking at the interface is ABA/BCB, thus S/Se first-neighbors are in *hollow positions*, and Se atoms eclipse Nb atoms, whereas S atoms are staggered with respect to W atoms – AB' corresponds to the bulk 3R stacking;

c) AB: the stacking at the interface is ABA/CBC, thus S/Se first-neighbors are in *hollow positions*, while both S-W and Se-Nb are staggered;

d) AA: the stacking at the interface is ABA/ABA, thus S/Se first-neighbors are on *top positions*, and Nb and W atoms are eclipsed;

e) A'B: the stacking at the interface is ABA/ACA, thus S/Se first-neighbors are on *top positions*, and Nb and W atoms are staggered.

In short, the major difference among these 5 interfaces is that AA', AB', and AB configurations have first-neighbor S/Se in *hollow stacking*, i.e., the energetically most stable contact, whereas AA and A'B have first-neighbors S/Se in *top stacking*: the latter stackings are introduced as models of the energetically less stable contacts induced by misorientation, i.e., a relative rotation, of the $NbS_2$ and $WSe_2$ systems in the synthesis process, as discussed in Ref.[ 25] and below.

To define the atomic coordinates we use experimental input from 2H bulk phases, taken from Ref. [36] for $NbS_2$ (a=3.33 Å, c=11.95 Å and S-S intralayer distance=2.97 Å) and from Ref. [37] for WSe$_2$ (a=3.282 Å, c=12.96 Å and Se-Se intralayer distance=3.34 Å), respectively. The lattice constant in the direction parallel to the interface was set to a=3.306 Å, as an average of NbS$_2$ 2H-bulk and WSe$_2$ 2H bulk lattice constants [36, 37]. The unit cell describing the NbS$_2$/WSe$_2$ vertical heterostructure (VH) is constructed by stacking 6 NbS$_2$ layers, followed by 12 WSe$_2$ layers and 6 NbS$_2$ layers, see Figure 1a, with the atomic configuration at the interface arranged according to the 5 stackings defined above. The minimum distance between S and Se atoms at the interface is the same for all the stackings, and is taken as the average of the S-S distance in bulk NbS$_2$ and Se-Se distances in bulk WSe$_2$.

Note that we use the experimental geometries for pure phases as input without relaxation in our DFT simulations to avoid artifacts due to a discrepancy between DFT+D3 predicted value and experimental one for the in-plane lattice parameter of NbS$_2$ (see Table S1 in the Supplementary Information). Additionally, in the absence of experimental data, we use an average value for the interlayer distance at the interface: this corresponds to a minimum in the total energy as predicted by our DFT+D3 approach, see Figure S2 in the Supplementary Information (where more information is also provided).

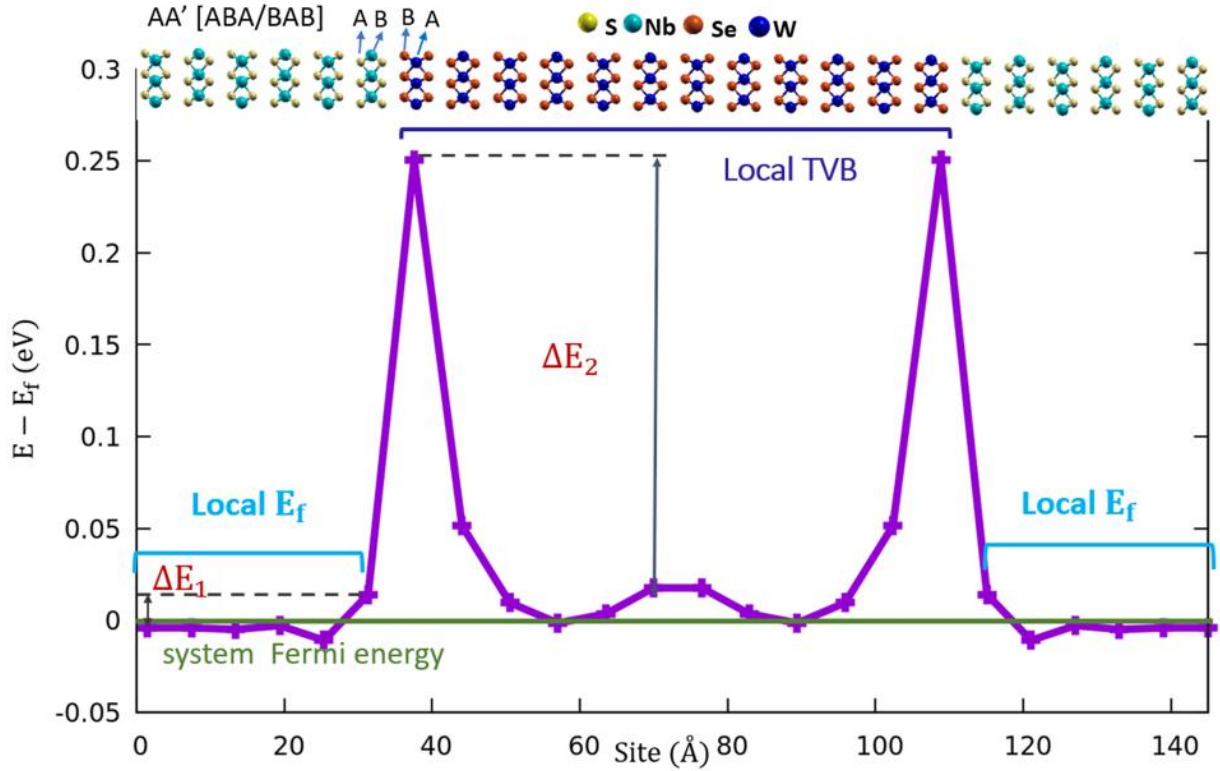

*Figure 2. Electrostatic potential profile of NbS2/WSe2 VH stack AA' with configuration ABA/BAB at the interface, $\Delta E_1$ is the difference between the local $E_f$ at the interface and lead (middle of NbS$_2$ part) where $\Delta E_2$ is the difference between the local TVB at the intertface and middle of WSe$_2$*

## 3. Results and Discussion

We performed DFT and transport simulations on the configurations built in Section 2, and complemented them with an analysis [21] based on partitioning a composite system and understanding its behavior in terms of constituent fragments, using the electrostatic potential as a unifying descriptor. Comparison with a previous study on the LH of the same $NbS_2/WSe_2$ materials combination [20] will allow us to assess the relative merits of the two interfaces.

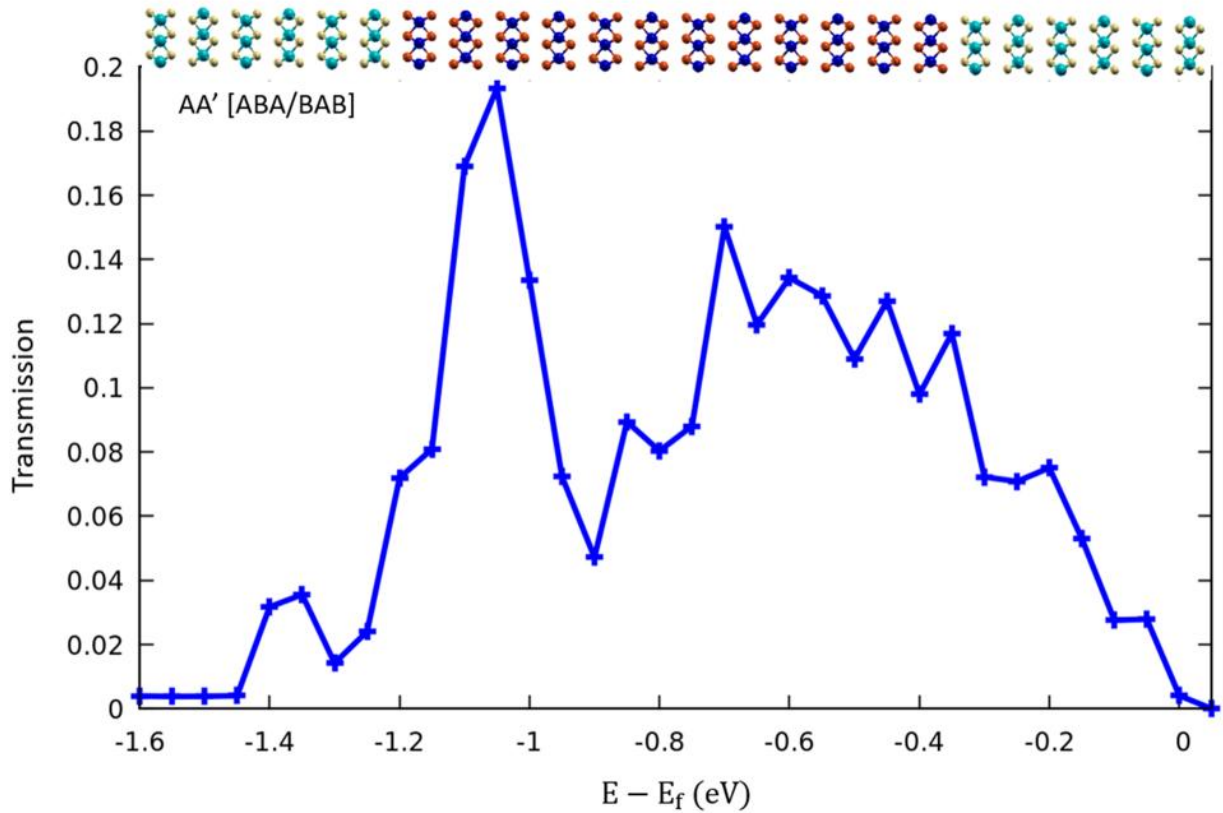

*Figure 3. Transmission curve for the NbS2/WSe2 VH with 2H stacking in the Bulk and AA' configuration at the interface.*

### 3.1 Electrostatic potential analysis

Focusing on the AA' stacking which corresponds to the lowest-energy (2H-like) interface, we conducted an electrostatic potential analysis [21] as illustrated in Figure 2.

In Figure 2 we plot the electrostatic potential on Nb and W atoms, shifted and aligned using quantities extracted from the corresponding fragment phases ($NbS_2$ and $WSe_2$), as discussed in Ref.[21] and applied to the $NbS_2/WSe_2$ LH in Ref.[20], so that the final plot corresponds to the local Fermi energy (loc-$E_F$) for the $NbS_2$ metallic fragment and the local top of the valence band (loc-TVB) for the $WSe_2$ semiconductor fragment. In practice, we build up a fragment 2H bulk $NbS_2$ phase by replicating the geometry in the middle of the $NbS_2$ phase, we extract from this calculation the difference between the Fermi energy ($E_F$) of the system and the electrostatic potential on the Nb atom, and we report this difference onto the $NbS_2/WSe_2$ VH by adding it to the values of the electrostatic potential on Nb atoms in the scattering region, thus finally determining the local Fermi energy $E_F$ (loc-$E_F$). Analogously, we build up a fragment 2H-bulk

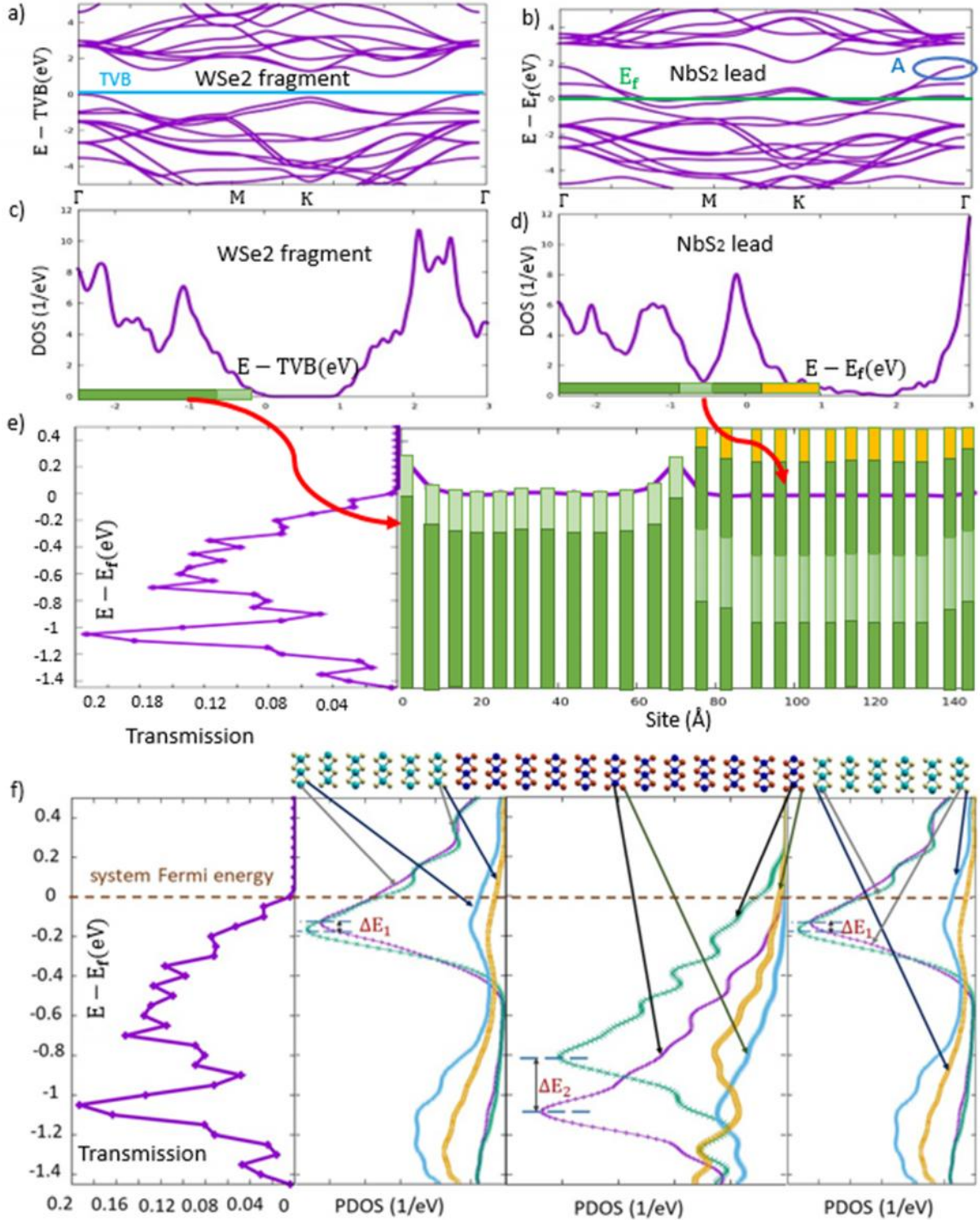

Figure 4. Transmission behavior analysis of NbS2/WSe2 VH (AA') structure: a) band structure of WSe2 fragment, b) band structure of NbS2 fragment, A is the TVB of NbS2  c) DOS of WSe2 fragment and d) DOS of NbS2 fragment (the fragments are taken from the *interfaces in the scattering region), e) transmission of AA' and band alignment procedure to find the states for electron transmission using the obtained color bar from DOS of the fragments in c) and d) in place of loc-E$_f$ and loc-TVB, and f) transmission of AA' and PDOS alignment to find the states for electron transmission. We choose atoms at the interface and far from the interface to do alignment using the PDOS of those atom. Note that $\Delta E_1$ is the shift between the PDOS of atoms from the middle and interface in NbS$_2$ part and $\Delta E_2$ is the shift between the PDOS of atoms from the middle and interface in part WSe$_2$.*

WSe$_2$ phase by replicating the geometry in the middle of the WSe$_2$ phase, we extract the difference between the TVB and the electrostatic potential on the W atom in this WSe$_2$ fragment, and we finally add this difference to the values of the electrostatic potential on W atoms in the scattering region to obtain the loc-TVB. Flatness of the loc-E$_F$ profile at the edges and of the loc-TVB in the middle of the unit cell guarantees convergence of the NbS$_2$ and WSe$_2$ fragments, i.e., guarantees that they are sufficiently far from (not affected by) the interfaces, and ensures precise projection of the wave function of the scattering system onto the wave function of the reference system in the successive transmission simulations. Incidentally, we observe that the concavity of the loc-TVB profile near the interfaces in Figure 2 indicates that the electronic density is transferred from WSe$_2$ to NbS$_2$ (see the discussion in Ref. [21]), which is consistent with band-structure alignment considerations. Note that we expect zero electric field inside the NbS$_2$ metallic phase so that the loc-E$_f$ should be flat, as it is indeed, whereas the profile of the loc-TVB is slightly more complex as a consequence of band bending and charge dipole effects at the metal/semiconductor interface.

The electrostatic potential analysis for the other stackings are reported and discussed in the Supplementary Information, Section 3. We note that the jump in the electrostatic potential at the interface in Figure 2 provides an estimate of the Schottky barrier for the NbS2/WSe2 metal/semiconductor VH interface, of the order 0.24 eV for the AA' configuration and changing slightly as a function of stacking (see Figure 2 and Figure S3). Our method qualitatively correlates with the CNL approach of Ref. [22], but the SBH estimate derived from the atomistic analysis comes out to be quantitatively smaller than the canonical estimate because the use of an atomistic NbS$_2$ TMD metal vs. an ideal metal (conductor) decreases the jump in the electrostatic potential at the interface. We finally add that one limitation of our approach is that we assume a rigid shift of the electronic states following the shift in the background potential, so that we expect that, when the electronic wave function at the interface is strongly deformed by large dipole moments, our method may become less accurate.

### 3.2 Transmission simulations and analysis

We performed transmission simulations for the configurations corresponding to all 5 considered stackings, always checking that the systems exhibit a flat potential (zero electric field) far from interfaces in the middle of the NbS$_2$ and WSe$_2$ regions, so that these regions can be used as leads in the transmission simulations (see Figure 2 and Figure S3). Since AA' is the most stable system, we start with a detailed analysis of its transmission first.

### 3.2.a The AA' stacking

Figure 3 reports transmission for the AA' stacking. Figure 3 is qualitatively similar to the analogous curve for the NbS$_2$/WSe$_2$ LH (Figure 3 in Ref.[20]), but with two important differences: (1) the range of finite transmission is much more extended (more than 1 eV here versus 0.2 eV in the LH), and (2) the curve shows a double peak structure with a minimum around -0.9 eV bias. Both these differences can be useful in applications, although the peaks are not located close to the Fermi level and Fermi level engineering would probably be necessary to use this feature in actual devices. Hereafter we demonstrate how they can be rationalized in simple terms via an electrostatic potential and fragment analysis.

In Figure 4 we analyze transmission via two complementary and consistent approaches.

First, we report in Figure 4a,b the band structure and in Figure 4c,d the DOS plots for the $NbS_2$ and $WSe_2$ fragment phases, showing that $WSe_2$ is a semiconductor and $NbS_2$ is a non-ideal metal exhibiting a pseudo-gap below the Fermi level. It can be noted that the DOS plots in Figure 4c,d exhibit intervals with high values of the DOS, highlighted with strong-green-colored bars, as well as intervals with intermediate to low values, highlighted with light-green-colored bars, and finally intervals with low DOS values, highlighted with yellow-colored bars. If we now draw these bars vertically and align them along the transmission direction by shifting their centers according to the electrostatic potential profile of Figure 2, we get the picture of Figure 4e. In other words, we align the colored bars by positioning them horizontally on the Nb and W atoms in the scattering region, and by shifting them vertically according to the values of the electrostatic potential on the corresponding atoms in the profile of Figure 2. Figure 4e quickly provides a pictorial count of the number of states available for transmission at any given value of the bias, thus allowing us to estimate the regions in which transmission is finite and the overall shape of transmission: a low number of available states corresponds to lower values of the transmission coefficient, while an increase (decrease) in the DOS is expected to bring about a peak (dip) in the transmission coefficient. The consistency of this approximate estimate with the actual transmission results reported for clarity on the left-hand-side of Figure 4e can be immediately appreciated. In more detail, transmission goes to zero above the Fermi energy because the TVB of $WSe_2$ is pinned at the Fermi energy $E_F$, and no states are available for electron transport above it (they fall in the band gap of $WSe_2$). Moreover, the drastic changes in DOS of the $NbS_2$ fragment around -0.65 eV to -1.2 eV are responsible for double peak behavior of transmission coefficient: the dip in transmission around -0.9 eV is understood as being caused by the corresponding dip in the $NbS_2$ DOS connected with its pseudo-band-gap.

In figure 4f we report PDOS plots, i.e., DOS projected onto individual atoms, that allow for an alternative analysis. As one can see in Figure 2, electrostatic potential jumps are largest for the atoms at the interface. We then focus on these interfacial atoms and contrast them with the atoms in the middle of each component phase ($WSe_2$ and $NbS_2$), including PDOS plots projected on both metal and chalcogenide atoms. First, these PDOS plots confirm the counting of states available for transmission from Figure 4e: the larger the PDOS, the larger the transmission around that energy (i.e., a peak in transmission), whereas a small PDOS for some of the atoms corresponds to a dip (valley) in the transmission coefficient. Second, we focus on the shift of the PDOS of Nb and W atoms between the middle of the corresponding fragment and the interface. We have highlighted these shifts as $\Delta E_1$ and $\Delta E_2$ in Figure 4f for Nb and W, respectively, and they read $\Delta E_1 = 0.03$ eV, $\Delta E_2 = 0.284$ eV, thus quantitatively coinciding with the corresponding differences in the atomic electrostatic potentials in Figure 2. The double-peak structure in transmission can thus be explained in terms of these two alternative and consistent analyses of PDOS shape and alignment. For comparison, we have also performed transmission simulations for pure $NbS_2$ (12 layers) and pure $WSe_2$ (12 layers), and report these results in Figure S4 of the Supplementary Information. As one can see from Figure S4, the transmission profile of the VH phase can be approximately interpreted as a convolution of the transmission plots of the pure phases, after proper alignment.

In a previous work [20] we have studied electronic transmission in the $NbS_2$/$WSe_2$ LH, that we also report in Figure S5 of the Supplementary Information for ease of comparison (see Figure 4 of Ref. [20]). Two major differences can be appreciated with respect to the present $NbS_2$/$WSe_2$ VH: first, a much wider energy interval of non-zero transmission; second, a double-peak structure in the VH case, absent in the LH. These differences can be rationalized. We note in fact that the LH in Ref. [20] is between monolayer (ML) systems, whereas the present VH involves bulk systems: this leads to distinct differences in the DOS and, in turn, in the transmission. A change in the number of layers from ML to bulk induces a change in the DOS and the band structure of the fragments (see Figure 4 and Figure S5), especially for the $NbS_2$ component: in the bulk system, at variance with the ML system, the DOS of the $NbS_2$ fragment does not have a gap below $E_F$, but rather a dip. A pseudo-gap is typically present in TMD (non-ideal) metallic layered materials below the Fermi energy, and in this respect $NbS_2$ is a paradigmatic example, but at the monolayer level the pseudo-gap becomes a real gap. We can thus explain the two major differences of the present results with respect to the $NbS_2$/$WSe_2$ LH [20]: the DOS of the $NbS_2$ fragment does not have a gap below $E_F$, but rather a dip around -0.7/-0.9 eV with respect to $E_F$. Correspondingly, the transmission coefficient does not go to zero below $E_F$, has a much wider range of finite values, but is reminiscent of the pseudo-gap (dip) by showing a double peak structure with a minimum around -0.9 eV. The change in the band structure when TMD goes from ML to multilayer form is in keeping with previous work ($NbS_2$[38], $WS_2$[39], ($MoS_2$, $WS_2$, $NbS_2$ and $ReS_2$)[40], ($MoS_2$,$WS_2$,$WSe_2$ and $MoSe_2$)[41]). These features of the $NbS_2$/$WSe_2$ VH may be advantageous in a possible use as a field effect transistor (FET), because the much broader range of finite transmission would make it more stable, while the double-peak shape could be exploited to obtain fine effects.

Clearly, the previous arguments are based on the relative alignment of the work function of $NbS_2$ and the TVB of $WSe_2$. It is thus important to validate our predicted values for these quantities against experiment and previous work. For the work function of $NbS_2$ we predict a value of 6.08 eV for the ML system (see also Ref. [20]) that is in good agreement with computational literature data [42], while we obtain 6.15 eV for the 2H bulk which is also in good agreement with previously reported data [18].

For the TVB of $WSe_2$ we obtain values of 5.038 eV for the ML system and 4.936 eV for the 2H bulk. These values are rather insensitive to the level of the DFT approach, and are basically coincident if one uses a gradient-corrected (GGA, semi-local) exchange-correlation (xc-) functional as we have done in the present work or uses a hybrid xc-functional [42]. The situation is less clear at the experimental level. Some experiments report values of the work function of $NbS_2$ much lower than 6 eV, around 4-5 eV (4 eV with an associated error of 1 eV in Ref.[ 43], 4.9 eV in Ref.[ 44]). However, these values were extracted from experiments conducted on highly defected $NbS_2$ samples, and it is to be expected that the work function of $NbS_2$ is very sensitive to defects, impurities, etc. More reliably, a $NbS_2/MoS_2$ system was investigated in Ref.[11]. The authors created thick stripes of $NbS_2$ and $MoS_2$ intersecting orthogonally and measured the work function of both system via Kelvin Probe Microscopy. It is important to underline that, under the conditions of the experimental set-up, the systems are highly n-doped, so that the measured values of the work function of these systems corresponds to the TVB of $NbS_2$ and the BCB of $MoS_2$, i.e., the valence band of $NbS_2$ is filled under the conditions of the experiment. The experimentally measured values are 4.81 eV for the TVB of $NbS_2$ and 4.53 eV for the BCB of $MoS_2$. The value of 4.53 eV for the BCB of $MoS_2$ is in keeping with previous literature, Ref.[45] and Ref.[ 46] (doped with $O_2$). The TVB of $NbS_2$ is predicted to be 5.2 eV at the PBE GGA level and 4.8 eV by the hybrid HSE xc-functional. The agreement between the $NbS_2$ TVB value measured in experiment and that predicted by the hybrid xc-functional is thus excellent. It can be noted that, whereas the lower region of the $NbS_2$ valence band and therefore the value of the work function are rather insensitive to the DFT approach, the top of the band comes out to be wider at the hybrid level, thus the larger discrepancy in the TVB value using GGA xc-functionals. However, the good match between the work function of $NbS_2$ via hybrid and GGA xc-functionals validates the accuracy of our DFT approach for the NbS2/WSe2 system.

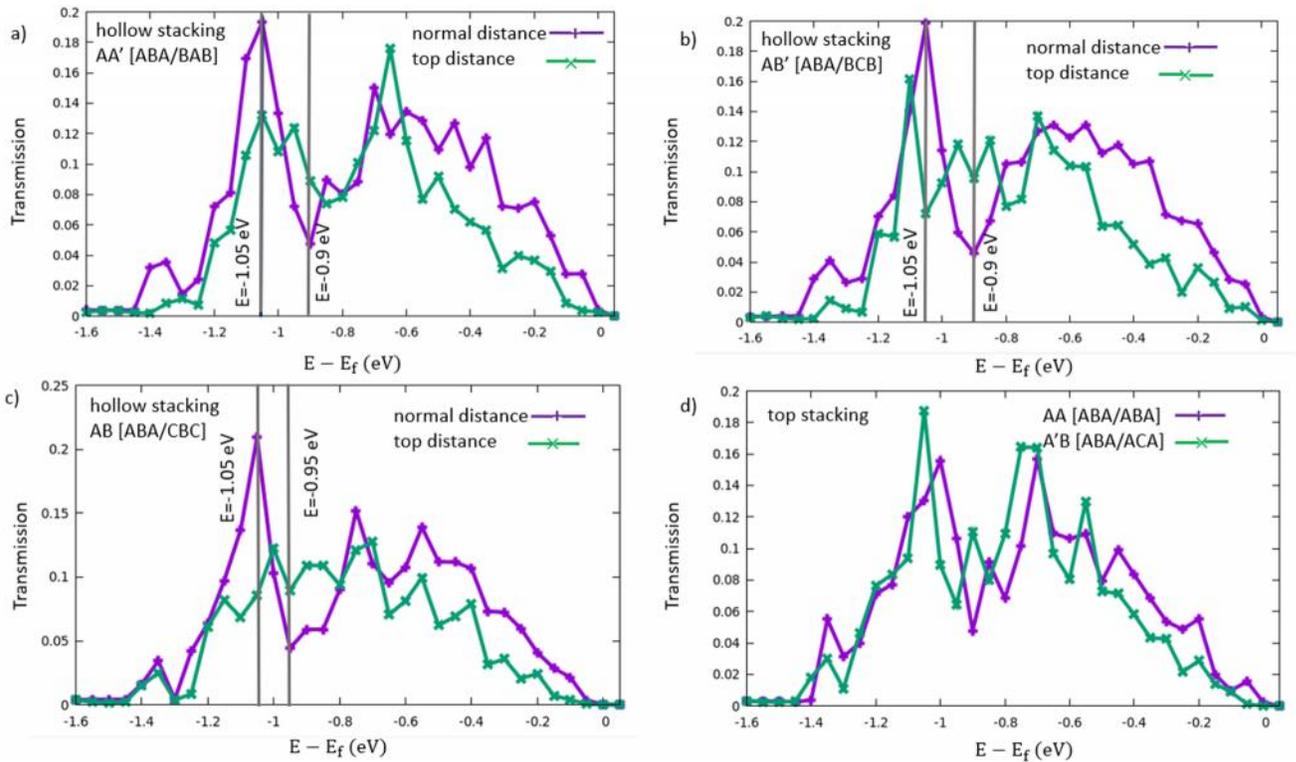

*Figure 5. Transmission coefficient of all stacking systems: a) hollow stacking AA' with normal distance and top distance at the interface, b) hollow stacking AB' with normal distance and top distance at the interface, c) hollow stacking AB with normal distance and top distance at the interface and d) top stacking AA and A'B*

### 3.2.a Other stacking epitaxies

We performed transmission simulations for all the five stacking geometries introduced in section 2 (see Figure 1). Moreover, to investigate the effects of a mis-orientation of the $NbS_2$ with respect to the $WSe_2$ phase, as in Ref.[25] we considered the three systems exhibiting hollow stacking, i.e., AA', AB' and AB, at two different values of the interfacial distance between S and Se atoms: a closest-approach distance between S and Se (normal distance), that we expect in the case of the energetically favorable epitaxy, and a "on-top" distance between S and Se atoms at the interface that corresponds to the interlayer spacing occurring for systems with on-top stacking, i.e., AA and A'B (top distance). The rationale for this choice is that a mis-orientation of the $NbS_2$ and $WSe_2$ fragments entails that in some region an on-top stacking will be realized, that will enforce an increase in the interlayer distance also for the hollow stackings in other regions of the VH [25]. We report the results of all these transmission simulations in Figure 5, illustrating the comparison between normal and on-top distances for the three systems with hollow stacking (Figures 5a-c), and separately the two systems with on-top stacking (Figures 5d). From Figure 5, we first observe that the transmission interval is similar for all stacking systems, so we predict that this feature will not be affected by a mis-orientation of the $NbS_2$ with respect to the $WSe_2$ phases. In contrast, the double-peak behavior is considerably lost when we increase the distances from normal to on-top for the three systems with hollow stacking. This phenomenon is most obvious for the AB stacking (Figure5c). To understand this finding in more depth, we focus on the AB phase here, while an analysis of the other two stackings is provided in the Supplementary Information, Section 4-3.

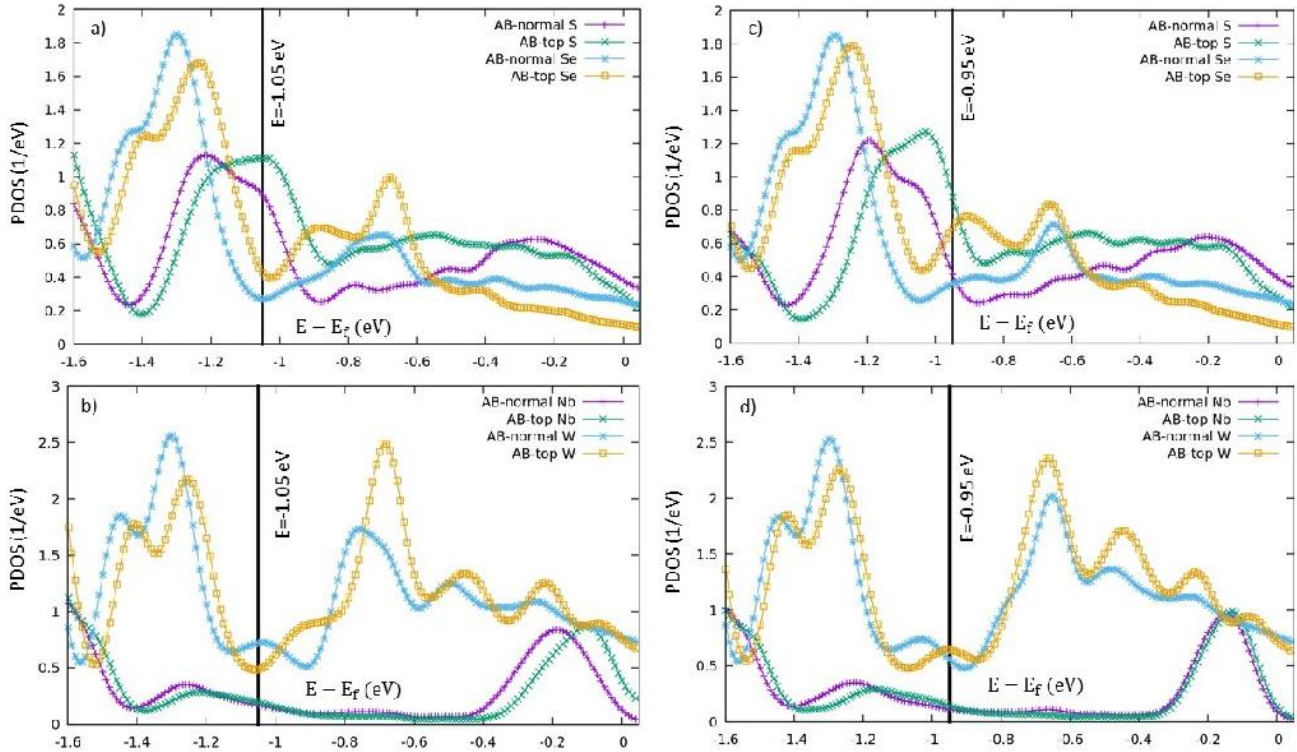

*Figure 6. PDOS plots of atoms at the interfaces of AB structure for both normal and top distances: a) PDOS of chalcogen atoms for the k-points with ninzero tranmission in E=-1.05 eV, b) PDOS of metal atoms for the k-points with nonzero transmission in E=-1.05 eV, c) PDOS of chalcogen atoms for the k-points with ninzero tranmission in E=-0.95 eV, d) PDOS of metal atoms for the k-points with nonzero transmission in E=-0.95 eV.*

We focus on two bias values with distinct changes in the transmission coefficient between normal and on-top distances, as highlighted in Figure 5: E = -0.95 eV, where a dip is found for normal distance, and -1.05 eV, where a maximum is found for normal distance. We then analyze the output of transmission simulations and single out the k-points which exhibit non-zero transmission at these two energies, thus dominating transmission. We then calculate and plot the PDOS of the interface atoms at these energies and at these k-points (the PDOS of the atoms far from the interface is not affected by the chosen interfacial distance, as shown in Figure S7), and report the corresponding plots in Figure 6. First, in Figure 6 we note a shift in the PDOS of the interfacial atoms to higher energies when the distance changes from normal to top. This however cannot be the reason of the change in the peak vs. dip in transmission because the changes in PDOS apparently do not correlate with the changes in transmission. For example, at E= -1.05 eV, at which bias we calculate a higher transmission coefficient for normal distance, the PDOS of most of the atoms at the interface is actually lower at normal distance with respect to on-top distance (only the PDOS of W at the interface is slightly higher at normal distance). A different phenomenon must then be responsible for the observed transmission behavior. We then hypothesize that *the lower transmission predicted for the top-distance systems is due to the fact that, when the distance increases, the atomic orbitals of the interfacial* atoms become more localized and the corresponding orbital overlaps decreases. To prove this, we performed a Wannier analysis of the wave function, determining the hopping elements from the Wannier90 Hamiltonian matrix [34, 35]. Then, we use a simple analytic formula and approximate transmission as the product of the PDOS and Wannier hopping matrix elements between the atoms at the interface. We estimate the current as [47]:

$$I = \frac{2e}{h} \int d\ T(E)[f_L(E) - f_R(E)] \quad (1)$$

where $T(E)$ is the transmission coefficient and $f_\alpha(E)$ is the Fermi-Dirac distribution function for lead $\alpha = L$ or $R$. In a simplified model with discrete energy levels, when we are in linear response regime, Eq.(1) can be written as [48]:

$$I = \frac{e}{\hbar} \int_{-\infty}^{+\infty} d\ D(E) \frac{(\gamma_1 \gamma_2)}{\gamma_1 + \gamma_2} [f_1(E) - f_2(E)] \quad (2)$$

where D(E) is the DOS at the energy E, and $\gamma_1$ ($\gamma_2$) is the coupling strength (hopping). Clearly, this approximation is quantitatively accurate only at low temperatures and modest values of the bias [48].

From Eqs. (1) and (2) we get approximately:

$$T(E) \approx D(E) \frac{(\gamma_1 \gamma_2)}{\gamma_1 + \gamma_2} \quad (3)$$

To apply Eq.(3) to our systems, we choose two paths for electron transmission, each consisting of two jumps (See Figure S8 in the Supplementary Information):

Path A: Nb    Se // Se    W

Path B: Nb    S // S    W

**Table 1.** Values of transmission along path A (via Se) and path B (via S) using the approximate formulae Eq.(4) at a bias E= -1.05 eV

| E = -1.05 eV | Transmission |
|---|---|
| Normal | 0.0585 (via Se) - 0.275 (via S) |
| Top | 0.0534 (via Se) - 0.1654 (via S) |

**Table 2.** Values of transmission along path A (via Se) and path B (via S) using the approximate formulae Eq.(4) at a bias E= -0.95 eV.

| E = -0.95 eV | Transmission |
|---|---|
| Normal | 0.0395 (via Se) - 0.0645 (via S) |
| Top | 0.0859 (via Se) - 0.1285 (via S) |

where the T(E) corresponding to each jump can be estimated as a product of the DOS on the involved Nb/W/S/Se atomic orbitals times the corresponding Wannier matrix elements among atomic orbitals, i.e.:

T(E) (path A via Se atom)    PDOS(Nb) PDOS(Se) PDOS(W) H(Nb,Se) H(W,Se)    (4a)

T(E) (path B via S atom)    PDOS(Nb) PDOS(S) PDOS(W) H(Nb,S) H (W,S)    (4b)

where H(a,b) is the matrix element of the Wannier90 Hamiltonian corresponding to the hopping between the orbitals of atoms a and b, PDOS(c) is the PDOS of an orbital corresponding to the given atom c, and to make the notation simpler we have not indicated explicitly the atomic orbitals, but a sum over all atomic orbital is implied in Eqs. (4a,b). In Table 1 and Table 2 we report the approximate values of transmission obtained using this simple formula at E= -1.05 eV and E = -0.95 eV, respectively. Notably, we find an *excellent agreement between the transmission values estimated via the approximate formula and the full transmission results* of Figure 5c: the transmission values via the dominating path B in the approximate formula closely correspond in ratio to those of the exact calculations.

It is worth noting in passing that the value of transmission in path B (via S) is higher than in path A (via Se) because of two reasons: first, the higher PDOS of S with respect to Se at the interface; second, the bigger hopping matrix element between W and S [H (W, S)] with respect to the hopping between Nb and Se [H (Nb, Se)]. Table S2 and Table S3 in the Supplementary Information, Section 4-2, report full details of these calculations.

## 4. Conclusions

In summary, we have investigated a metal ($NbS_2$)/semiconductor ($WSe_2$) VH via DFT and transmission simulations. We consider individual fragments in the VH to be in 2H bulk phases and we change the geometry at the interface according to all 5 possible epitaxial stackings. We perform band structure QM simulations ensuring convergence of periodic models and investigate the transport properties of the VHs. We find that the transmission profile as a function of bias exhibits a wide finite interval and a *double-peak structure* for all 5 stackings when at contact distance. We then analyze the electrostatic potential profile and estimate the transmission behavior either using band alignment of the separated fragments, or using PDOS in the scattering region, obtaining results consistent between the two approaches and in agreement, and therefore fully rationalizing, the exact simulations. In contrast, the transmission profile flattens when the distance between the chalcogen atoms at the interfaces increases from contact (normal) to on-top distances (simulating of mis-orientation of the layers at the interface). Importantly, in this case band alignment arguments are not sufficient to explain this behavior. We thus introduce a simple transmission model that includes Wannier transfer matrix elements at the interface. This extended model now allows us to rationalize quantitatively our findings. Interestingly, the interval of nonzero transmission coefficient for the $NbS_2$/$WSe_2$ VH is qualitatively similar but much wider with respect to the previously investigated LH [20]. This, together with two-peak structure in transmission, makes this system potentially interesting for applications in electronic devices.

**Competing Interests**

No competing interests were disclosed.

**Grant Information**

This project has received funding from the European Union's Horizon 2020 research and innovation programme under grant agreement No. 829035 (QUEFORMAL, https://www.queformal.eu/), and from the MIUR PRIN Five2D project under grant agreement n. 2017SRYEJH.

**Acknowledgments**

The ISCRA programme of the CINECA supercomputing Center (LHFET-BS project) is gratefully acknowledged for computational resources.

**Supplementary Information**

The Supplementary information is available on the journal website, and includes additional information about Modeling, Structure analysis, Electrostatic potential profile analysis and transmission analysis. Figures S1-S10 refer to: electrostatic potential energy profile and transmission for two different geometries of $NbS_2$/$WSe_2$ VH, the energy of the system as a function of interface distance, the electrostatic potential profile for the stackings, transmission analysis of NbS$_2$/WSe$_2$ VH using transmission of pure phases, the transmission analysis of NbS$_2$/WSe$_2$ LH, comparision between electrostatic potential energy profile of the atoms in AB stacking in top and normal distance, PDOS of the chalcogen atoms of AB structure far from the interfaces, the schematic representation of defined paths for the electron jumping at the interface of AB structure, Transmission coefficient analysis of AA' and AB', respectively. Table S1 compares the result of DFT+D3 calculation with experimental data obtained for NbS$_2$ and WSe$_2$ 2H-bulk unit cell. Also, Table S2 and Table S3 have the numbers for the parameters in Eq.4.

# Supplementary information for:

# Vertical Heterostructures between Transition-Metal Dichalcogenides - A Theoretical Analysis of the NbS$_2$/WSe$_2$ junction

Zahra Golsanamlou[1], Poonam Kumari[1], Luca Sementa[1,*], Teresa Cusati[2], Giuseppe Iannaccone[2], Alessandro Fortunelli[1,*]

[1] CNR-ICCOM and IPCF, Consiglio Nazionale delle Ricerche, via G. Moruzzi 1, Pisa 56124, Italy

[2] Dipartimento di Ingegneria dell'Informazione, Università di Pisa, Via G. Caruso 16, Pisa 56122, Italy

* Corresponding authors: luca.sementa@cnr.it, alessandro.fortunelli@cnr.it

## 1) Model

### 1-1) More details about the transmission calculation using PWCOND

We performed transmission simulations using the PWCOND routine within the QE package [1,2], based on a scattering-state approach which integrates numerically a scattering equation in real space along the direction of transport according to the formula: $T = \sum_m |T_m|^2 = \text{Tr}[\mathbf{T}^+\mathbf{T}]$ where, m (n) is related to the propagating and decaying states on the left (right) leads and **T** is the matrix of normalized transmission amplitudes, $T_m = \sqrt{\frac{I_m}{I_n}} t_m$. Here, $I_m$ and $I_n$ are the currents carried by the state m and n, respectively. This approach employs scattering theory for rightward propagating modes from the left to the right electrode [1], holds in the limit of ballistic transport and corresponds to ideal transmission in the absence of defects.

### 1-2) Wannier90

The Wannier representation of the system electronic structure in terms of localized orbitals is a representation based on real-space single-particle wave functions defined through unitary transformation of the Bloch orbitals and labelled according to the band index n and the lattice translation vector R:

$$w_n(\mathbf{r}) = \frac{V}{(2\pi)^3} \int_B \left[\sum_m U_m^k \psi_m(R)\right] e^{-i\mathbf{j}\cdot\mathbf{R}} d \qquad (S1)$$

where V represents the volume of the unit cell, and $U^k$ is a unitary transformation which mixes the Bloch states. The choice of $U^k$ is not unique. Different choices can lead to different Wannier functions which will have different spatial extensions. This non-uniqueness of $U^k$ is due to the fact that the orbitals represented by Bloch states belong to a set of bands that are separated by energy gaps but have degeneracies within them o that at each k point there will be many unitary transformations possible. This may make the use of Wannier functions unsuitable in the case of realistic (realistically complicated) problems. A procedure to eliminate this arbitrariness was proposed by Marzari and Vanderbilt in Ref. [3]. In this method, iteratively redefined transformations lead to a uniquely defined set of maximally localized Wannier functions (MLWFs). This approach can be applied to a variety of problems starting from isolated systems to periodic solids and is coded in the last release of the Wannier90 code [3, 4].

**2) Structure analysis**

We constructed $NbS_2$-2H bulk based on experimental data extracted from two different sources: Ref. [5] and Ref. [6]. We obtained 3.567 Å for interlayer distance (S-S distance) using the parameters in Ref. [5] and 3.358 Å using the parameters in Ref. [6]. The interlayer distance obtained based on Ref. [5] is much closer to the $MoS_2$ interlayer distance (3.49 Å) obtained in independent experimental work [7] so we decided to continue with the geometrical parameters reported in Ref. [5] (named structure1 below). Indeed, Figure S1a and S1b show the potential profile and transmission coefficient, respectively, for the structures taken from Ref. [5] (structure1) and from Ref. [6] (structure2). It is apparent from Figure S1a that the electrostatic potential profile on atoms in structure 2 are less converged as compared to structure 1. Also, the energies of the $WSe_2$ region is not well converged for structure 2 since the flat potential on the two W atoms in the middle of $WSe_2$ region is farther from the Fermi energy. Finally, the transmission coefficient calculated on structure 2 is typically smaller than that calculated on structure 1 as shown in Figure S1b, confirming that structure2 [6] is less realistic.

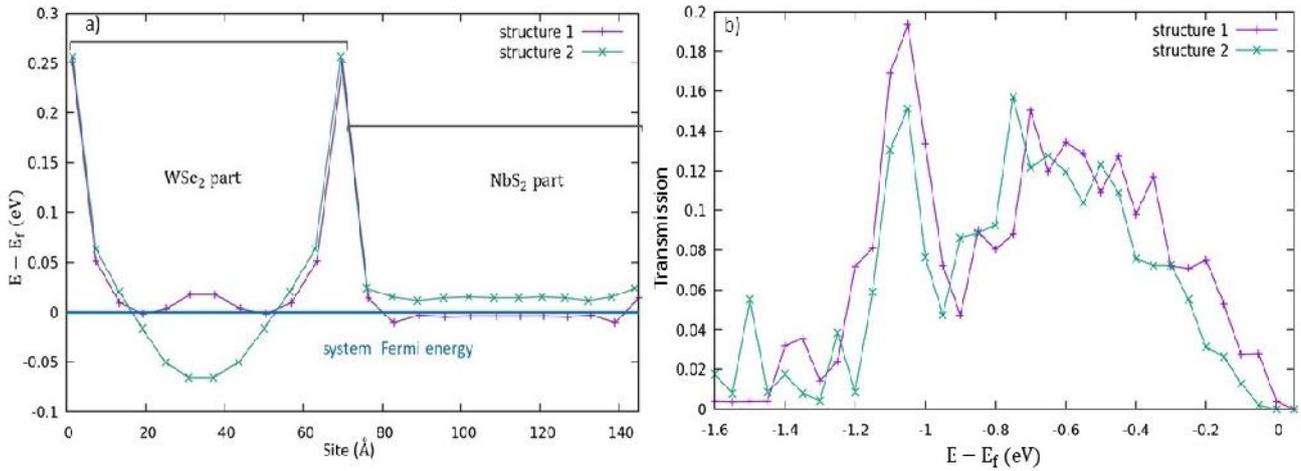

***Figure S1.*** *a) electrostatic potential energy profile and b) transmission coefficient of $NbS_2/WSe_2$ VH based on two different geometries reported for $NbS_2$ 2H-bulk. Structure1 was taken from Ref. [5] and structure 2 was taken from Ref. [6].*

As mentioned in the main text we use the experimental data to construct the geometries of the pure phases in the VH to avoid artifacts due to a discrepancy between DFT+D3 predicted value and experimental one for the in-plane lattice parameter in NbS$_2$. Table S1 reports a comparison between the distances of NbS$_2$ and WSe$_2$ 2H-bulk unit cell as obtained from experiment, Ref.[5] for NbS$_2$ and Ref.[7] for WSe$_2$, respectively, and DFT+D3.

**Table S1.** Comparison between the experimental and theoretical data for 2H-bulk unit cells of NbS$_2$ and WSe$_2$

|  | DFT+D3 | | | Experimental data | | |
| --- | --- | --- | --- | --- | --- | --- |
|  | $d_{X-X}$ (in-plane) | $d_{X-X}$ inter layer | Lattice parameter | $d_{X-X}$ (in-plabe) | $d_{X-X}$ inter layer | Lattice parameter |
| NbS$_2$ | 3.12 Å | 3.44 Å | 3.44 Å | 2.97 Å | 3.56 Å | 3.33 Å |
| WSe$_2$ | 3.39 Å | 3.64 Å | 3.33 Å | 3.34 Å | 3.66 Å | 3.28 Å |

*X=chalcogen atom

Additionally, for the AA′ VH we calculated the total energy of a system in which the geometries of the pure phases were frozen as discussed above, but using different S-Se distances at the interface, around the average value between S-S and Se-Se distances (i.e., 3.617  ). One can see from Figure S2 that the minimum energy is obtained at the the average value between S-S and Se-Se distances (3.617  ), so that we used this values for all VH stackings.

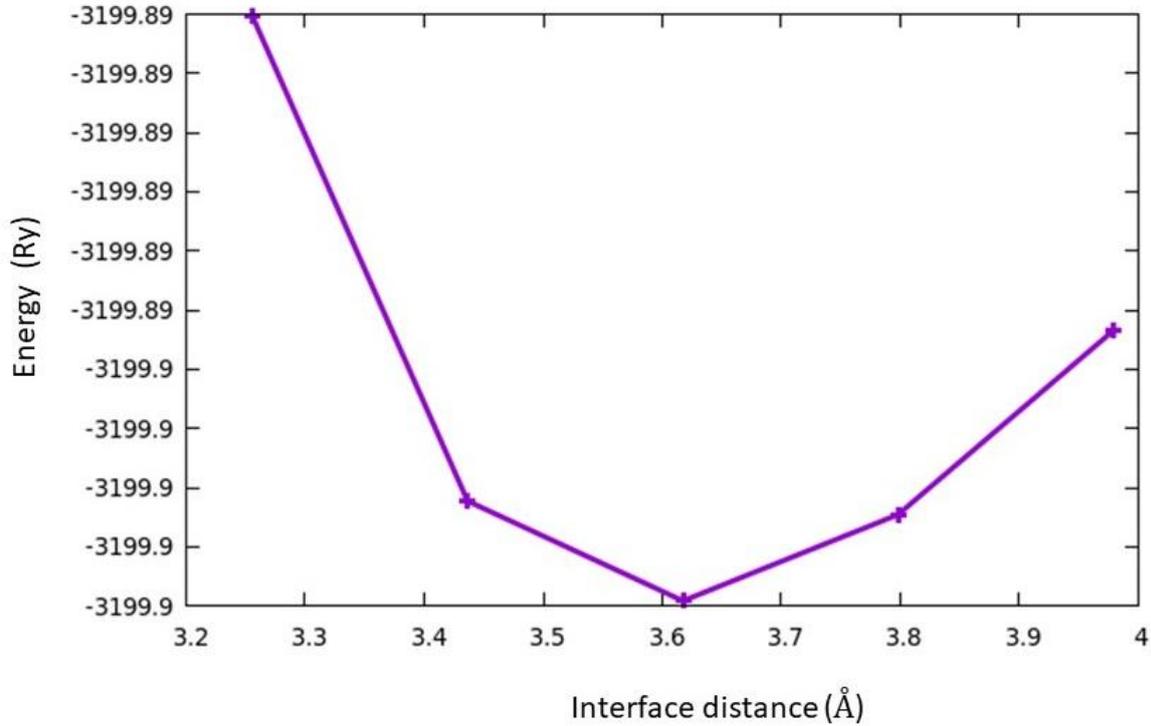

*Figure S2. The energy of the AA' VH for different distances at the interface,*

**3) Electrostatic potential profile analysis for stacking systems**

We consider five stackings for $NbS_2/WSe_2$ VH where the atomic configurations of the fragments are in the 2H-bulk phases and the atomic configurations at the interface change according to the definition of stackings in the main text, Section 2 and also Figure 1.

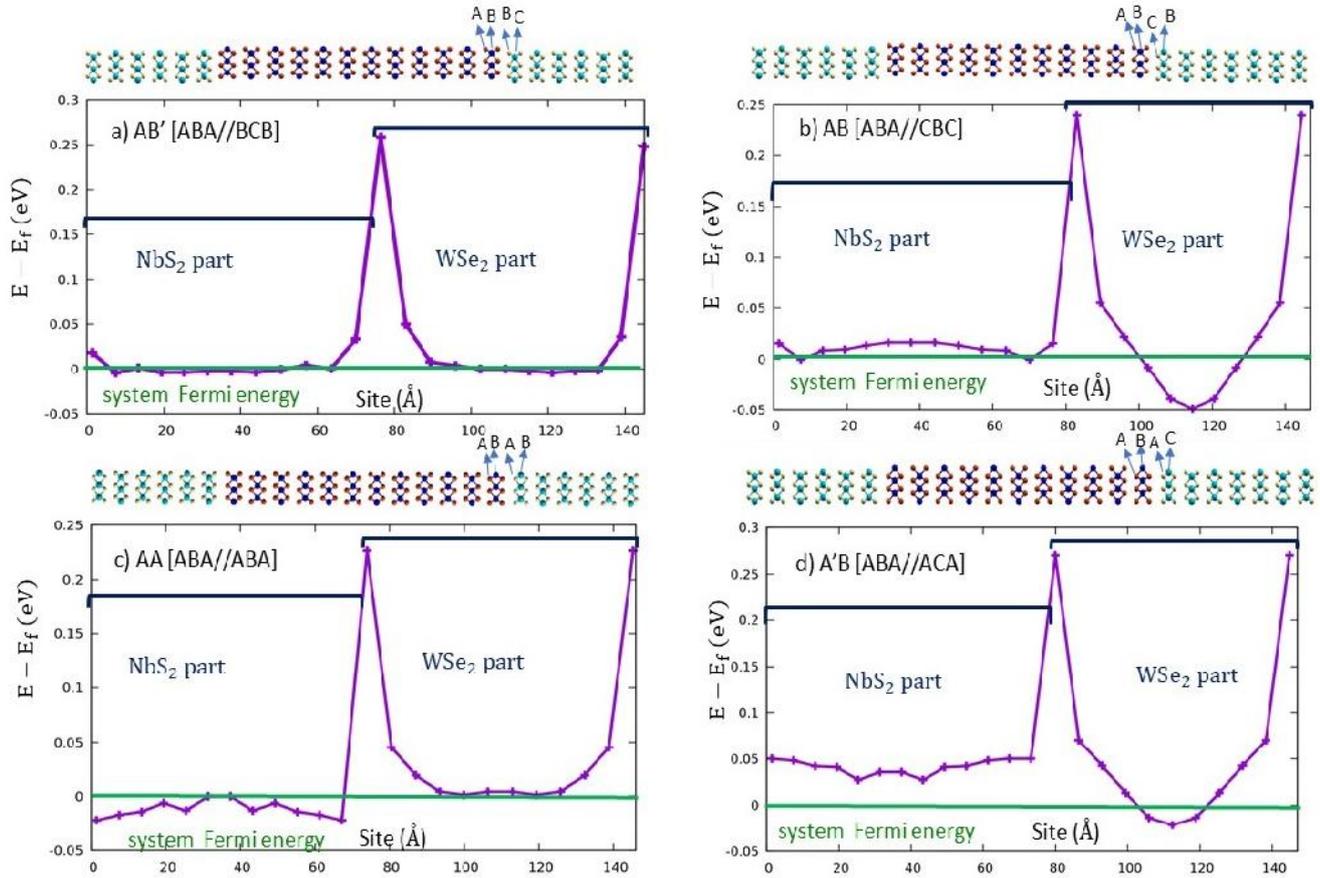

*Figure S3.* The electrostatic potential profile for a) AB' [ABA/BCB], b) AB [ABA/CBC], c) AA [ABA/ABA] and d) A'B [ABA/ACA].

We apply our electrostatic potential analysis [8,9], for all stacked geometries not reported in the main text, as depicted in Figure S3. Apparently, the changes in the geometry at the interface affect the electrostatic potential on the atoms. Note that in AA', AB' and AB, chalcogen atoms are in hollow positions at the interface whereas in AA and A'B the chalcogen atoms are in top positions at the interface. We obtain a converged behavior of the potential profile in both $NbS_2$ and $WSe_2$ regions of the AB' and AA structures (Figure S5a and Figure S5c). The convergence of local TVB on the W atoms in the AB and A'B structures is instead a bit slower and exhibit more deviation from the $E_f$ of the system. This is simply due to the lower number of $WSe_2$ layers in these structures where we transform one $WSe_2$ layer into $NbS_2$ layer for reasons of symmetry. For all the jumps at the potential profile, the dipole effects at the interfaces are compensated. Overall, we can observe that potential profile of all the stacking systems are converged enough to perform transmission simulations: the oscillations in the $NbS_2$ part are small (in order of 0.01 eV) allowing us to choose this region (the middle of $NbS_2$ part) as the leads.

## 4) Transmission analysis

### 4-1) Transmission analysis of NbS2/WSe2 VH using the perfect transmission of the leads

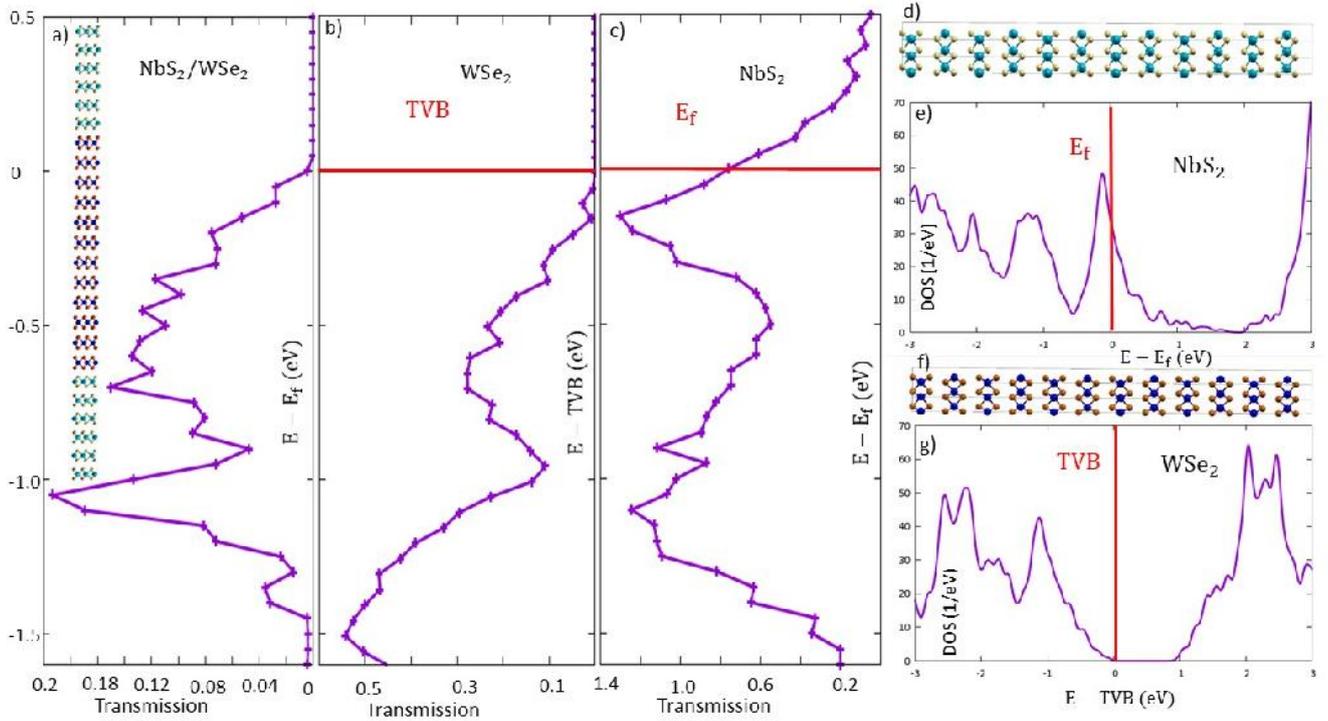

**Figure S4.** Transmission behavior analysis of *NbS$_2$/WSe$_2$ AA' VH using pure phases: a) transmission spectra of NbS$_2$/WSe$_2$ AA' VH b) transmission spectra of pure WSe$_2$ in 2H-bulk phase, c) transmission spectra of pure NbS$_2$ in 2H-bulk phase d) NbS$_2$ 2H-bulk structure used for the calculation, e) density of states (DOS) of pure NbS$_2$, f) WSe$_2$ 2H-bulk structure used for the calculation g) DOS of pure WSe$_2$*

We can analyze the transmission behavior of the NbS$_2$/WSe$_2$ AA' VH in terms of the transmission of pure phases, as illustrated in Figure 5. By comparing Figures S5a-S5c with Figures S5d-S5e, we can observe that transmission of the composite system is zero when there is a gap in the transmission spectra of one of the pure phases, which in turn is associated with a gap in the DOS of the phase. Analogously, transmission in the VH increases when transmissions of both pure phases increase, or decreases when there is a reduction in transmissions of pure phases, whose origin lies in the electronic properties of the pure phases, see Figure S5e and S5g.

### 4-2) Transmission analysis of NbS2/WSe2 lateral heterostructure

We report here the transmission analysis for the NbS$_2$/WSe$_2$ lateral heterostructure (LH) and compare with the results for the NbS$_2$/WSe$_2$ VH from the main text.

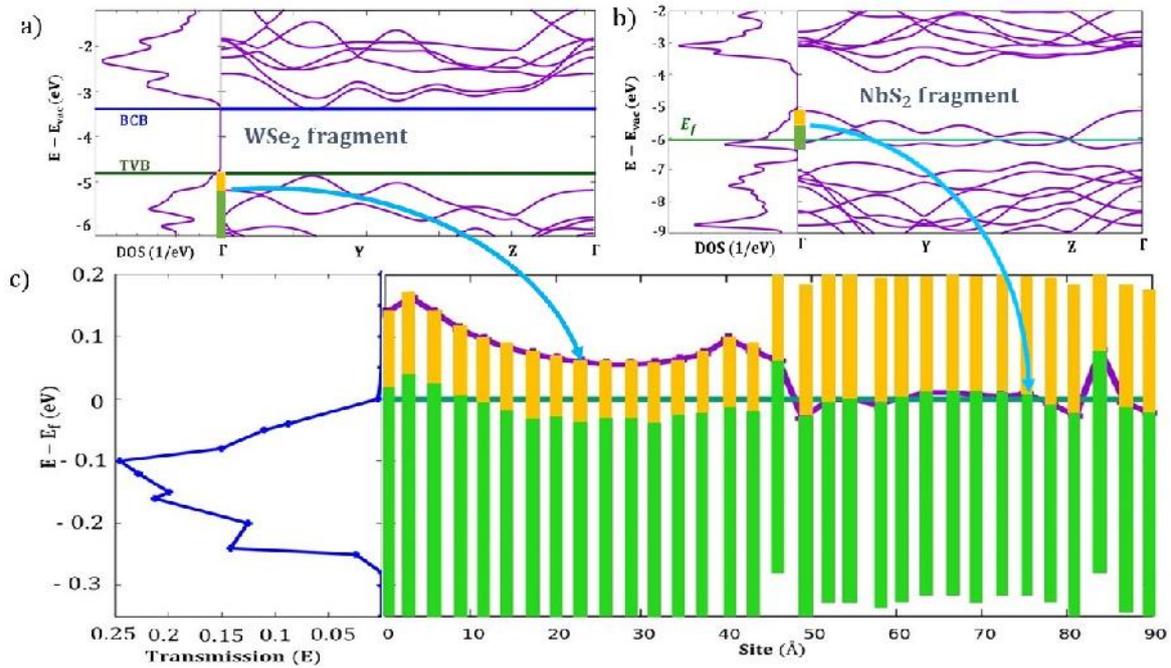

*Figure S5. Our approach to find the nonzero energy interval for transmission analysis: a) DOS and band structure of $WSe_2$ fragment, b) DOS and band structure of $NbS_2$ fragment, c) transmission of $WSe_2/NbS_2$ LH and band alignment procedure to find the states for electron transmission using the obtain color bar from a) and b) in place of local $E_f$ and TVB. This figure is taken from* Adv. Theory Simul. **2020**, 2000164. DOI: https://doi.org/10.1002/adts.202000164.

In Figure S5 we define color bars indicating regions with lower density of states (DOS) in yellow and regions with higher DOS in green where the DOS of the fragments is calculated on geometries taken by replicating the scattering region of $NbS_2/WSe_2$ LH. Then we put these DOS color bars along the transmission direction in the positions of Nb and W atoms by shifting them up and down in the energy scale according to the electrostatic potential energy profile of Nb and W atoms. This allows us to provide a good estimate of the interval of nonzero transmission coefficient based on the number of states available for electron transmission.

**4-2) Further analysis of the transmission behavior of the AB structure**

Figure S6 represents the electrostatic potential energy profile on the atoms in the AB structure at normal and top distances. Our analysis shows that increasing interatomic distances at the interface affects the electrostatic potential on the atoms by a small shift. The largest changes are observed for the S and Se atoms located at the interfaces (consistent with the behavior of the PDOS, see Figure 4 in the main text).

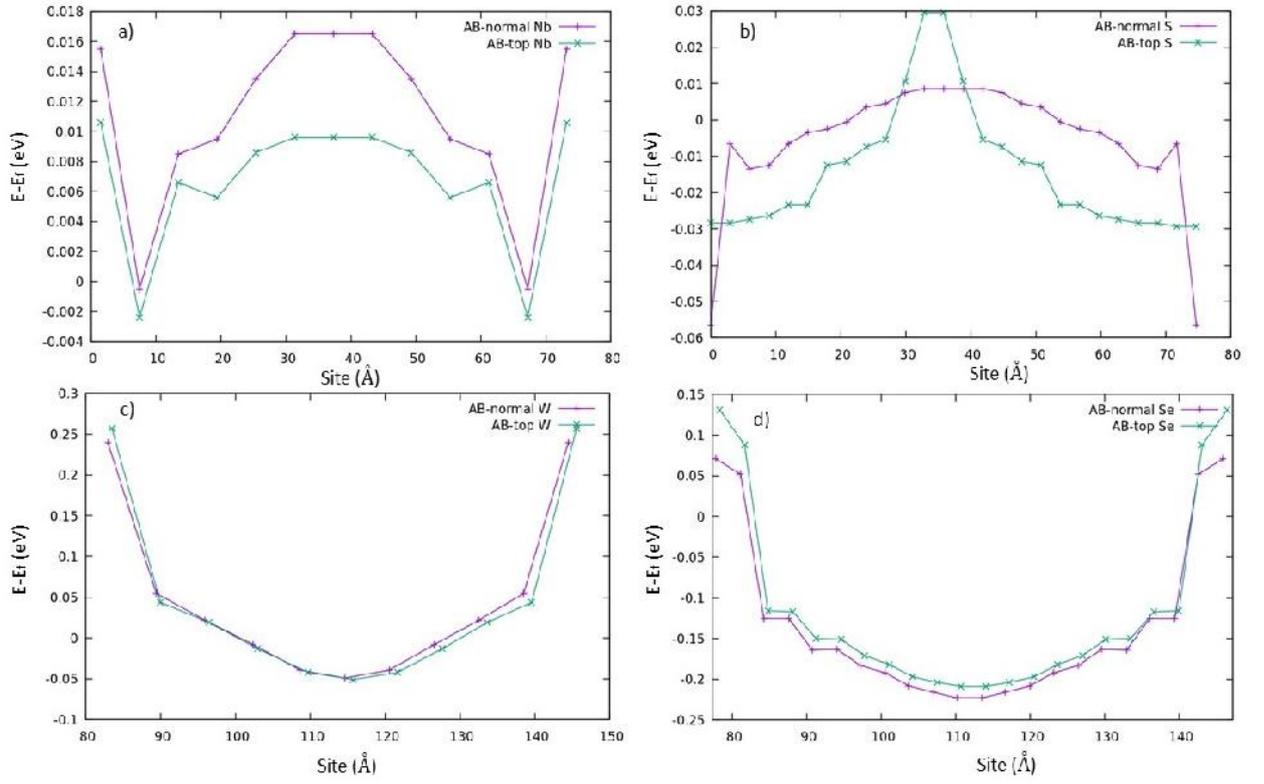

***Figure S6.*** *Comparision between electrostatic potential energy profile of the atoms in scattering region of AB in normal and top distances: a) Nb atoms, b) S atoms, c) W atoms and d) Se atoms*

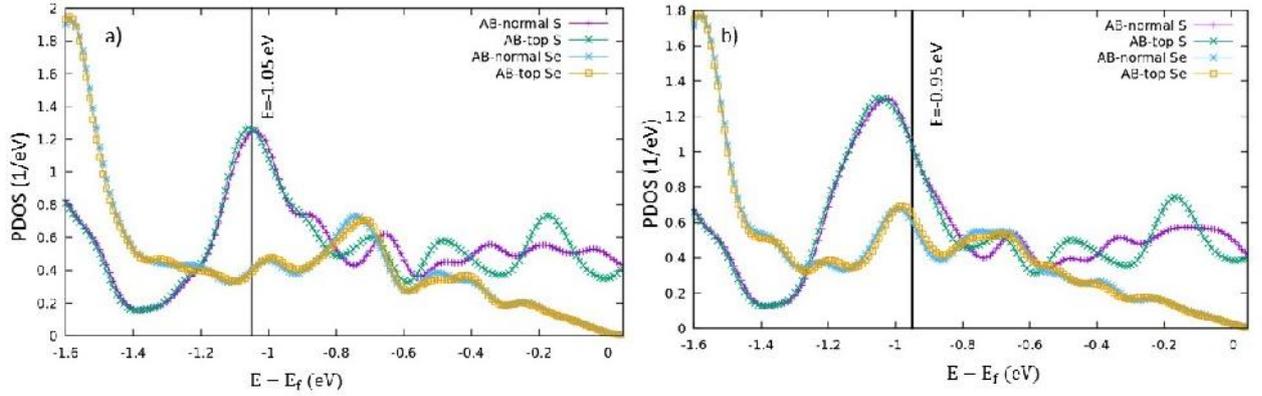

***Figure S7.*** *PDOS of the chalcogen atoms in the AB structure far from the interface: a) for the k-points with nonzero transmission for bias E=-1.05 eV and b) for the k-points with nonzero transmission for bias E=-0.95 eV.*

Figure S7 shows that the atoms in the AB structure far from interfaces have the same PDOS at most bias values. In Figure S7 we report the PDOS for the k-points which have nonzero transmission at representative energies (E=-1.05 eV and E=-0.95 eV), we see that the PDOS at these energies are the same for the atoms far from the interfaces, so we focus on the PDOS of the atoms at the interface in the main text and indicate in Figure S6.

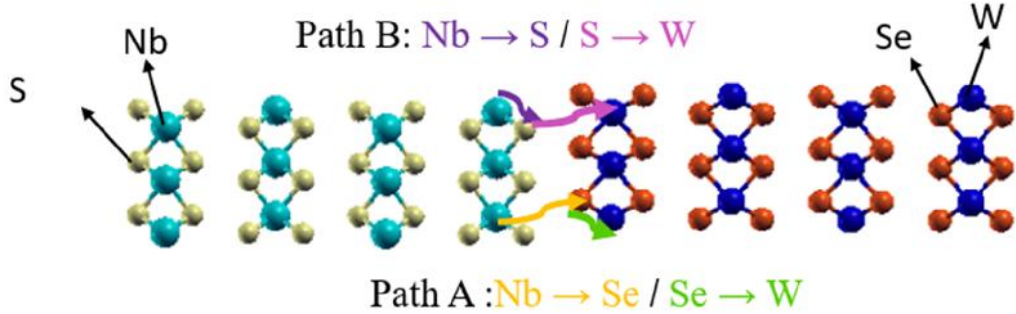

*Figure S8.* *Schematic representation of the defined paths for the electron jumps at the interface of AB structure*

**Table S2.** The numerical values of the quantities used in the approximate transmission formula (Eq.4a and 4b) introduced in the main text at E=-1.05 eV.

| AB [ABA/CBC] Max of Transmission (E=-1.05 eV) | |
|---|---|
| Normal | Top |
| PDOS(Nb) = 0.168 (1/eV) | PDOS(Nb) = 0.193 (1/eV) |
| PDOS(S) = 0.891 (1/eV) | PDOS(S) = 1.111 (1/eV) |
| PDOS (W) = 0.723 (1/eV) | PDOS (W) = 0.485 (1/eV) |
| PDOS(Se) = 0.267 (1/eV) | PDOS(Se)= 0.424 (1/eV) |
| H (W, S) = 0.331 eV | H (W, S) = 0.205 eV |
| H (W, Se) = 7.478 eV | H (W, Se) = 7.590 eV |
| H (Nb, S) = 7.691eV | H (Nb, S) = 7.779 eV |
| H (Nb, Se) = 0.241 eV | H (Nb, Se) = 0.177 eV |

**Table S3.** The numerical values of the quantities used in the approximate transmission formula (Eq.4a and 4b) introduced in the main text at E=-0.95 eV.

| AB [ABA/CBC] Min of Transmission (E=-0.95 eV) | |
|---|---|
| Normal | Top |
| PDOS(Nb)= 0.110 (1/eV) | PDOS(Nb)= 0.142 (1/eV) |
| PDOS (S)= 0.414 (1/eV) | PDOS (S)= 0.880 (1/eV) |
| PDOS (W)= 0.557 (1/eV) | PDOS (W)= 0.646 (1/eV) |
| PDOS(Se)= 0.358 (1/eV) | PDOS(Se)= 0.696 (1/eV) |
| H (W, Se) = 7.478 eV | H (W, Se) = 7.590 eV |
| H (W, S) = 0.331 eV | H (W, S) = 0.205 eV |
| H (Nb, S) = 7.691 eV | H (Nb, S) = 7.779 eV |
| H (Nb, Se) = 0.241 eV | H (Nb, Se) = 0.177 eV |

Also, the PDOS analysis shows that at the interface there is a charge transfer from $WSe_2$ to $NbS_2$ (the PDOS of S is larger than Se) which decreases when we increase the interlayer distance at the interface. This is consistent with the curvature of the electrostatic potential profile In Figure S6c and S6d.

## 4-3) Analysis of the transmission behavior of AA' and AB' structures as a function of the interlayer distance at the interface

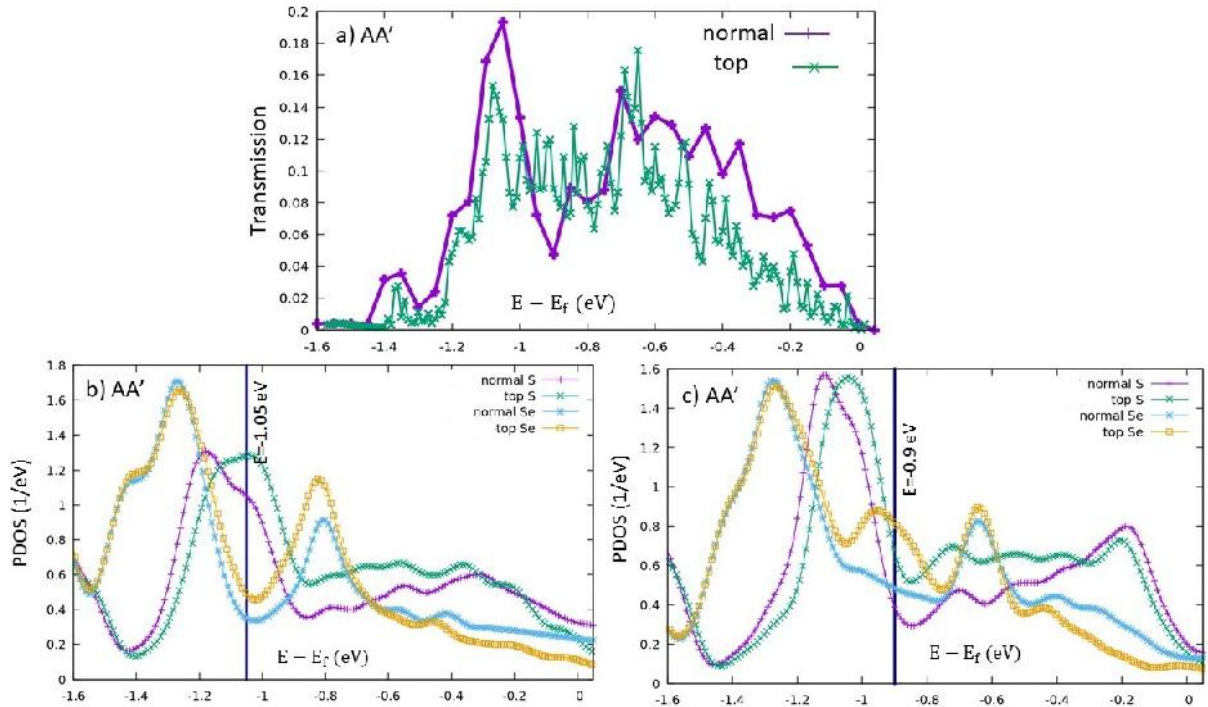

*Figure S9.* a) Transmission coefficient of AA' structure in normal and top distances, b) partial density of state (PDOS) of S and Se atoms at the interfaces for the k-points which produce nonzero transmission at E=-1.05 eV, c) partial density of state (PDOS) of S and Se atoms at the interfaces for the k-points which produce nonzero transmission at E=-0.9 eV.

In the main text, Figure 5, we report the transmission coefficient for all the stacking systems at normal distance while, in Figure 5a and Figure 5b, we also report the transmission for hollow structures (AA', AB' and AB) when the distance at the interface changes from normal to top distance. We find a double peak structure for the transmission of all stacking systems in Figure 5 at normal distance, but when we increase the distance at the interface from normal to top the double peak behavior is lost. This effect is more distinct for the AB structure, so we study the changes in transmission for the AB structure in the main text, while here we provide additional information on the AA' and AB' structures. Figure S6a and S9a shows the transmission coefficient at normal and top distances for AA' and AB', respectively. We select two values of bias at which peak and dip of transmission occur at normal distance. Then we calculate the PDOS for the k-points which have nonzero transmission at those values of bias. From Figure S9b, S9c, S10b and S10c one can see the shifts in the PDOS of the S and Se atoms which explain the changes in the transmission coefficient. Also the PDOS plots in Figure S9c and S10c show the bigger overlap for atomic orbitals of top distance at E= -0.9 eV which proves higher transmission value for this configuration. Using the approximate formula for transmission, equations 4a and 4b in the main text, we then predict similarly a higher value for transmission at normal distance for E=-1.05 eV and a higher value for transmission at top distance for E=-0.9 eV.

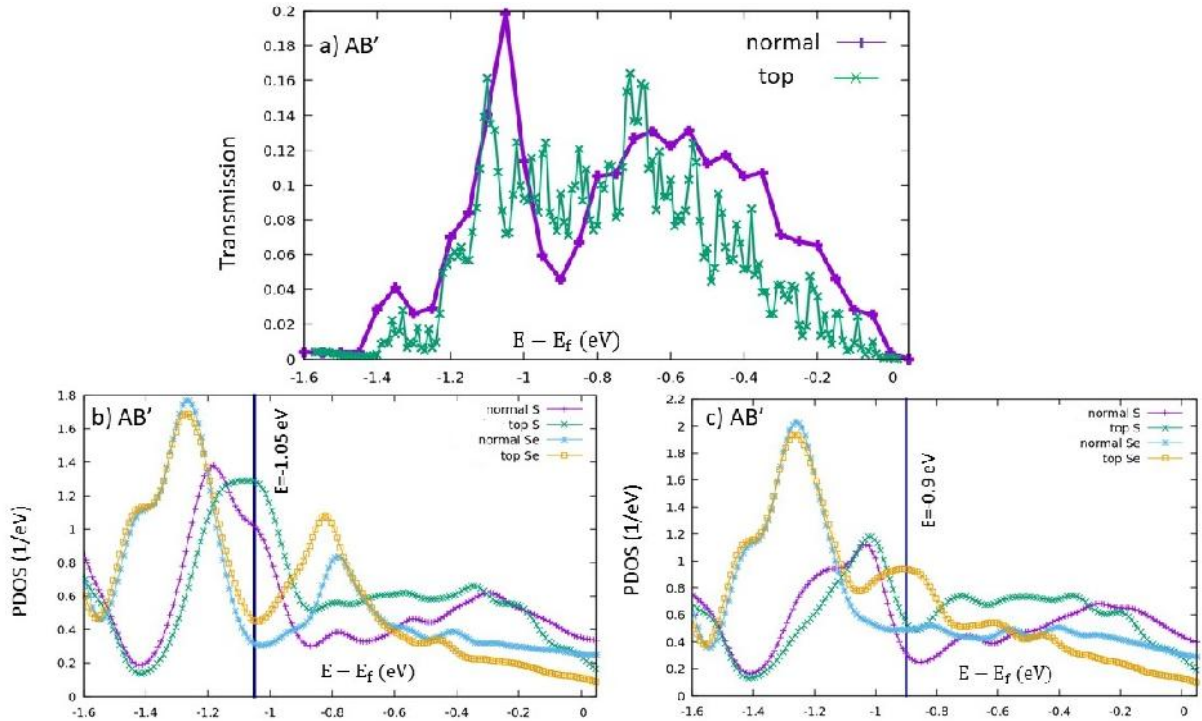

*Figure S10.* a) Transmission coefficient of AB' structure in normal and top distances, b) partial density of state (PDOS) of S and Se atoms at the interfaces for the k-points which produce nonzero transmission at E=-1.05 eV, c) partial density of state (PDOS) of S and Se atoms at the interfaces for the k-points which produce nonzero transmission at E=-0.9 eV.